\documentclass[twocolumn,showpacs,preprintnumbers,amsmath,amssymb,floatfix]{revtex4}
\usepackage{epsf}

\begin{document}

\newcommand{\tilr}{{\tilde r}}
\newcommand{\hA}{{\hat{A}}}
\newcommand{\hB}{{\hat{B}}}
\newcommand{\hC}{{\hat{C}}}

\title{Interior of Nonuniform Black Strings}
\author{
{\bf Burkhard Kleihaus}
}
\affiliation{{ZARM, Universit\"at Bremen, Am Fallturm, 
D--28359 Bremen, Germany}
}
\author{
{\bf Jutta Kunz}
}
\affiliation{
{Institut f\"ur Physik, Universit\"at Oldenburg, Postfach 2503,
D--26111 Oldenburg, Germany}
}
\date{\today}

\begin{abstract}
We consider nonuniform black strings inside their event horizon.
We present numerical evidence, that the singularity touches the
horizon as the horizon topology changing transition is reached.
\end{abstract}

\maketitle

{\sl Introduction --}
Already Kaluza and Klein invoked a fifth dimension
to unify gravity and electromagnetism, 
hiding the higher dimension subsequently by rendering it small and compact.
Following these ideas, today 
supergravity and string theory also put forward the existence of
higher dimensions in their attempt to unify the fundamental forces of nature.

In such $D$-dimensional manifolds with $p$ compact dimensions,
black holes can either be localized in the compact dimensions,
or the black hole horizon can wrap the compact dimensions completely.
Black holes which are localized in the compact dimensions
have the horizon topology of a $(D-2)$-sphere, $S^{D-2}$,
and are called `caged' black holes \cite{Kol:2004ww,Harmark:2007md}.
In contrast, when the horizon wraps the compact dimensions,
the horizon topology reflects the topology of the compact manifold.
In the simplest case, the single compact dimension is simply a circle, $S^1$.
The black holes then have the horizon topology of a torus, 
$S^{D-3} \times S^1$,
and are referred to as black strings \cite{Kol:2004ww,Harmark:2007md}.

As long as caged black holes are much smaller than the compact dimension,
they resemble more or less $D$-dimensional Schwarzschild black holes
(when they are static and uncharged).
But as their size and with it their mass grows,
they begin to feel the finite size of the compact dimension 
and deform accordingly,
reaching a maximal size beyond which they no longer fit
into the compact dimension
\cite{Kudoh:2003ki,Kudoh:2004hs}.
Black strings, on the other hand, exist for all values of the mass.
However, as shown by Gregory and Laflamme (GL) \cite{Gregory:1993vy},
such `uniform' black strings (UBS), 
which do not depend on the compact coordinate,
become unstable below a critical value of the mass.
GL therefore suggested, that unstable UBS
would decay to black holes, which possess higher entropy.

In contrast, Horowitz and Maeda~\cite{Horowitz:2001cz} argued 
that the horizon of UBS
could not pinch off in finite affine time. 
They therefore conjectured that the solutions would settle down 
to nontranslationally invariant solutions with the same horizon
topology as the original configurations, `nonuniform' black strings (NUBS).
Shortly after, such NUBS were found perturbatively
\cite{Gubser:2001ac} and numerically
\cite{Wiseman:2002zc}. 
However, the NUBS cannot serve as endpoints of the instability, 
since they are too massive 
and possess too low an entropy \cite{Gubser:2001ac}, 
at least as long as the number of dimensions is smaller than 13 
\cite{Sorkin:2004qq},
indicating that black holes are the end-state of the instability, nevertheless
\cite{Kol:2004ww}.

Emerging from the UBS branch at the marginally stable solution,
the NUBS form a branch of solutions along which the deformation 
of the horizon along the compact direction perpertually increases. 
This is illustrated in Fig.~1, where we exhibit
the spatial embedding of the horizon of static $D=6$ black string solutions
for a sequence of NUBS, yielding a geometrical view of the 
increasing nonuniformity of the solutions \cite{Kleihaus:2006ee}.
We here parametrize the deformation by 
$\lambda= \frac{1}{2} \left( \frac{{R}_{\rm max}}{{R}_{\rm min}}
 -1 \right)$ 
\cite{Gubser:2001ac},
where ${R}_{\rm max}$ and ${R}_{\rm min}$
represent the maximum and minimum radius of a $(D-3)$-sphere on the horizon.

\begin{figure}[h!]
\begin{center}
\epsfysize=5cm
\mbox{\epsffile{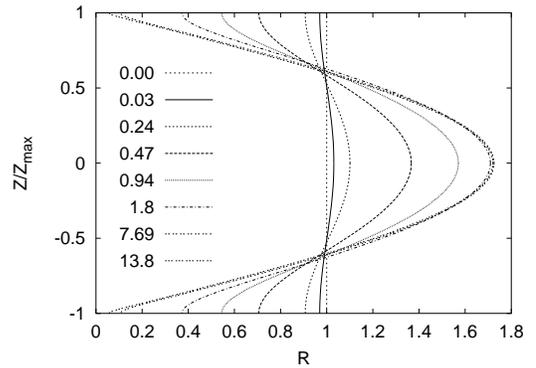}}
\caption{
Spatial embedding of the horizon of $D=6$ black strings
with increasing deformation $\lambda$,
ranging from $\lambda=0$ to $\lambda=13.8$.
($Z$ denotes the proper length along the compact direction,
$R$ the proper radius of the horizon.)
}
\end{center}
\label{f-1}
\end{figure}

Extrapolating the numerical results to $\lambda \rightarrow \infty$ shows
that the maximal radius ${R}_{\rm max}$ assumes a finite limiting value, while
the radius $R_{\rm min}$ of the `waist' of the black strings shrinks to zero
\cite{Wiseman:2002zc,Kudoh:2004hs,Kleihaus:2006ee}.
In the limit $\lambda \rightarrow \infty$,
the horizon will then pinch off, changing the horizon topology from 
$S^{D-3}\times S^1$ to $S^{D-2}$
at a singular `merger' configuration
\cite{Kol:2002xz,Wiseman:2002ti,Kol:2003ja,Kol:2004ww}.
Study of the global and horizon properties of NUBS and black holes
in $D=5$ and $D=6$ dimensions indeed provides persuasive evidence that
the NUBS branch merges with the black hole branch at such a
topology changing transition \cite{Kudoh:2004hs,Kleihaus:2006ee}.

A deep question associated with the envisaged transition
is whether it is associated with the occurrence of a naked singularity
and thus a violation of cosmic censorship.
Indeed, it appears possible, that as the black string pinches,
a naked singularity is formed, because the singularity, which orginally
winds the compact dimension, gets broken \cite{Kol:2004ww}.
To address this question,
we here investigate the inside of nonuniform black strings,
to see how the curvature and in particular the singularity
evolve, as $\lambda$ becomes large.
Our results indicate, that the singularity will touch the horizon
at the topology changing transition.

{\sl Nonuniform Black Strings --}
We consider the Einstein action 
\begin{equation}
I = \frac{1}{16\pi G} \int_M d^Dx \sqrt{-g}R 
   - \frac{1}{8\pi G} \int_{\partial M} d^{D-1}x \sqrt{-h}K \ , 
\label{action}
\end{equation}
for a $D$-dimensional spacetime with one compact direction. We denote the 
periodic length of the compact direction by $L$.
Here the last term is the Gibbons-Hawking surface term 
\cite{Gibbons:1976ue}.

A convenient form of the metric is given by \cite{Kleihaus:2006ee}
\begin{equation}
ds^2 = -e^{2A} f dt^2 +e^{2B}\left(\frac{d\tilr^2}{f}+dz^2\right)
      + e^{2C}\tilr^2 d\Omega_{D-3}^2 \ , 
\label{metric1}
\end{equation}
where $z$ is the coordinate of the compact direction. 
$A$, $B$, and $C$ are functions of $\tilr$ and $z$ only,
and $f= 1-(\tilr_0/\tilr)^{D-4}$. 
Thus the horizon resides at $\tilr = \tilr_0$.
The functions $A$, $B$, and $C$ have to be determined 
from the Einstein equations 
$G_t^t = 0$, $G_\tilr^\tilr+G_z^z = 0$ and $G_\theta^\theta = 0$.
The remaining nontrivial equations $G_\tilr^\tilr-G_z^z = 0$ 
and $G_z^\tilr = 0$ form the constraints.
(For $A=B=C=0$ UBS arise.)

We now focus on black strings in $D=6$ dimensions, since
they can be determined with great numerical accuracy 
up to large deformations \cite{Kleihaus:2006ee}.
For solutions outside the horizon, $\tilr \geq \tilr_0$, 
we introduce the coordinate $\bar{r}$ via 
$\tilr = \sqrt{\tilr_0^2 + \bar{r}^2}$, 
while inside the horizon, $\tilr \leq \tilr_0$, 
we define the coordinate $r$ via $\tilr = \sqrt{\tilr_0^2 - r^2}$. 
In these coordinates the horizon resides at $\bar{r}=0=r$. 
Substitution in the line element
Eq.~(\ref{metric1}) then yields for $\tilr \leq \tilr_0$
\begin{equation}
ds^2 = e^{2\hA} r^2 dt^2 +  e^{2\hB} (-dr^2+dz^2) 
    +  e^{2\hC} d\Omega_{3}^2 \ , 
\label{metric2}
\end{equation}
where $e^{2\hA}= e^{2A}/(\tilr_0^2-r^2)$, $e^{2\hB}= e^{2B}$,
and $e^{2\hC}= e^{2C}(\tilr_0^2-r^2)$.
Inside the horizon the Einstein equations are hyperbolic, 
the coordinate $r$ playing the role of `time',
\begin{eqnarray}
\hA,_{rr} & = & \hA,_{zz}-\frac{\hA,_r}{r} -(\hA,_r+\frac{1}{r})(\hA,_r+3 \hC,_r) 
\nonumber \\
& &	       
               +\hA,_z(\hA,_z+3 \hC,_z) \ , 
\nonumber\\
\hB,_{rr} & = & \hB,_{zz}+3 \hC,_r (\hA,_r+\frac{1}{r}+ \hC,_r)-3 \hC,_r (\hA,_z+\hC,_z) 
\nonumber \\
& &	       
               +3 e^{2(\hB-\hC)} \ , 
\nonumber\\
\hC,_{rr} & = & \hC,_{zz}-\hC,_r (\hA,_r+\frac{1}{r}+ 3 \hC,_r) +\hC,_z (\hA,_z+3 \hC,_z) 
\nonumber \\
& &	       
               -2 e^{2(\hB-\hC)} \ , 
\label{PDEs}	       
\end{eqnarray}
Initial conditions are given at the horizon (setting $\tilr_0 = 1$)
\begin{eqnarray}
& &
\hA(0,z) = A_{\rm H}(z) , \ 
\hB(0,z) = B_{\rm H}(z) , \ 
\hC(0,z) = C_{\rm H}(z) , 
\nonumber\\
& &
\partial_r\hA|_{r=0} =\partial_{\bar{r}} A|_{\bar{r}=0} = 0 \ , \ 
\partial_r\hB|_{r=0} =\partial_{\bar{r}} B|_{\bar{r}=0} = 0 \ , \ 
\nonumber\\
& &
\partial_r\hC|_{r=0} =\partial_{\bar{r}} C|_{\bar{r}=0} = 0 \ , \ 
\label{initialC}
\end{eqnarray}
where the functions $A_{\rm H}(z)$, $B_{\rm H}(z)$ and $C_{\rm H}(z)$ 
are obtained from the solution outside the horizon.
The origin of the $z$ coordinate is chosen such that
${R}_{\rm min} = e^{C_{\rm H}(0)}$ and ${R}_{\rm max} = e^{C_{\rm H}(L/2)}$.

Setting 
$G_t^t =G_r^r+G_z^z=G_{\theta_1}^{\theta_1}=G_{\theta_2}^{\theta_2}
=G_{\varphi}^{\varphi}= 0$ in the identities 
$\nabla_\mu G^{\mu r}= 0$ and $\nabla_\mu G^{\mu z}= 0$  reveals that the constraints
satisfy the advection equations
$$
\left(\partial_r \pm \partial_z\right)
\left( \sqrt{-g}(G^r_z \mp (G^r_r-G^z_z)/2)\right) = 0 \ .
$$
Consequently, the constraints vanish everywhere, 
since they vanish at the horizon.

{\sl Numerics --}
The hyperbolic differential equations (\ref{PDEs}) are re-written
as a set of first order equations in $r$. For the evolution in $r$ we use 
a fourth order Runge-Kutta method \cite{footnote}.
The `spatial' coordinate is scaled via $z \rightarrow z/L$.
The difference formulae for 
$z$ derivatives are obtained
from a polynomial approximation of the functions:
If a function $y$ has values $y_k$ at the gridpoints $z_k$, we approximate the 
$q$th derivative of $y$ at $z_k$ by
\begin{equation}
y^{(q)}(z_k) = \sum_{j=k-n}^{k+n} y_j P^{(q)}_{j,k}(z_k) \ .
\label{difz1}
\end{equation}
Here $P^{(q)}_{j,k}(z)$ is the $q$th derivative of the Lagrange polynomial
of degree $ 2 n$, 
$$ 
P_{j,k}(z)= \prod_{\begin{array}{c}{\scriptstyle l=k-n}\\ {\scriptstyle l\neq j}\end{array}}^{k+n}
        \frac{z-z_l}{z_j-z_l} \ .
$$ 
To maintain the same order of approximation at all gridpoints we
include auxiliary gridpoints $\{-z_{n},-z_{n-1}, \cdots ,-z_2\}$
and $\{1+z_1,1+z_2, \cdots ,1+z_n\}$.
The values of the functions at these points are obtained 
from the symmetry properties $y(-z)=y(z)$ and $y(1+z)=y(1-z)$. 

However, if a singularity is encountered at some point $z_s$, 
we have to restrict to the interval $ z_s < z < 1-z_s$. 
In this case the approximation of the $z$ derivatives is less `symmetric'. 
Thus, if $z_{k_s}$ is the gridpoint next to $z_s$, and $k_s\leq k<k_s+n$, 
we define
\begin{equation}
y^{(q)}(z_k) = \sum_{j=k_s}^{k_s+2n} y_j P^{(q)}_{j,k}(z_k) \ , \ 
P_{j,k}(z)= \prod_{\begin{array}{c}{\scriptstyle l=k_s}
 \\ {\scriptstyle l\neq j}\end{array}}^{k_s+2n}
        \frac{z-z_l}{z_j-z_l} \ ,
\label{difz2}
\end{equation}
and similarly for gridpoints close to $1-z_s$.

As we argue below, the solution becomes singular if $e^\hC$ tends to zero.
In order to avoid the singularity we have to restrict to a domain,
where $e^\hC$ is larger than some small number, typically $\approx 10^{-6}$.
Thus the computation is performed in two stages. 
In the first stage the solution evolves from the horizon ($r=0$) 
until it becomes nearly singular at $z=0$ and $z=1$ for some $r=r_s$. 
During this stage we employ the difference formula (\ref{difz1}).
In the second stage the solution evolves from  $r=r_s$. 
In each Runge-Kutta step we check 
whether the solution is nearly singular at gridpoints $z_{k_s}$. 
If this is the case, 
we restrict to  $z_{k_s}< z_k < 1-z_{k_s}$ in the next Runge-Kutta step.
During this stage the difference formula (\ref{difz2}) is employed.
The evolution stops when the $z$-interval shrinks to zero.

Typical step sizes used are $\Delta r \approx 1 - 4 \times 10^{-4}$ 
and the number of gridpoints ranges between $N=100$ and $N=2400$. 
The order of the difference formula is $2n=6-14$.
We have checked the consistency of the solutions 
for different choices of $\Delta r$, $N$ and $n$, 
and also for different nonequidistant gridpoint distributions.

{\sl Results --}
We have constructed NUBS solutions inside their horizon for 
several values of the deformation parameter $\lambda$. 
Near the horizon the solutions extend 
smoothly into the interior region. 
The function $e^\hC$ decreases monotonically for fixed $z$.
When $e^\hC$ tends to zero at some $r_s(z)$, 
the Kretschmann scalar diverges there,
indicating that the curvature singularity resides at $r_s(z)$.  
For UBS the radial coordinate of the singularity is constant,
$r_s(z)=1$,
while for NUBS with small deformation 
$r_s(z)$ shows an oscillation about $r=1$ with small amplitude, 
which then increases with increasing $\lambda$.
We exhibit the coordinate $z_s(r)$ of the singularity 
in Fig.~2 for several values of the deformation parameter $\lambda$.
(Note the similarity with Fig.~1.)

\begin{figure}[h!]
\begin{center}
\epsfysize=5cm
\mbox{\epsffile{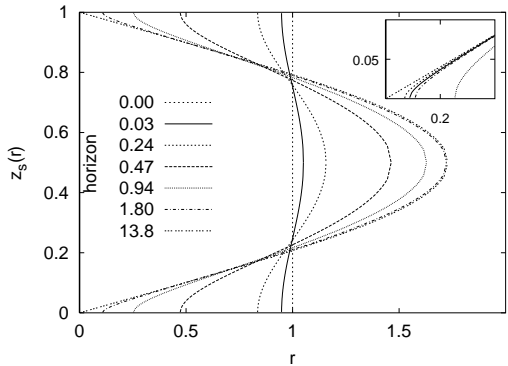}}
\caption{The coordinate $z_s(r)$ of the location of the singularity 
is shown for several values of the deformation parameter $\lambda$
of black strings, ranging from $\lambda=0$ to $\lambda=13.8$.
}
\end{center}
\label{fig1}
\end{figure}

As seen in the figure, 
the singularity approaches the horizon at $z=0$ and $z=1$,
when $\lambda$ becomes large.
This is further demonstrated in Fig.~3,
where we exhibit the minimal distance 
$R_{\rm sing} = \int_{r_{\rm H}}^{r_s} e^{\hat B(r,0)} dr $ 
of the singularity from the horizon at the waist.
We note that for large deformation the distance decreases approximately
inversely proportional to the deformation, $R_{\rm sing} \sim \lambda^{-1}$.
Thus we extrapolate that in the limit $\lambda \rightarrow \infty$
the singularity will touch the horizon at the waist of the string,
i.e., a naked singularity will appear
at the topology changing transition \cite{Wiseman:2002ti,Kol:2004ww}.

\begin{figure}[h!]
\begin{center}
\epsfysize=5cm
\mbox{\epsffile{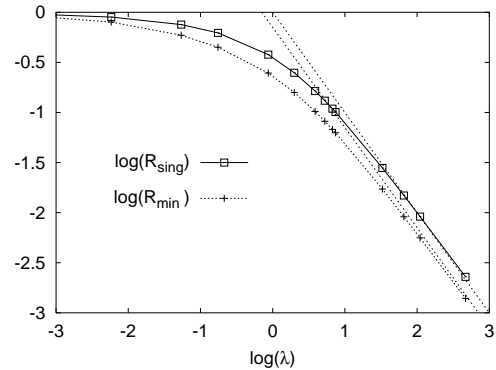}}
\caption{ 
The distance $R_{\rm sing}$ from the horizon
to the singularity at the waist ($z=0$)
is shown versus the deformation $\lambda$
for a sequence of NUBS, 
together with its approximation at large $\lambda$,
$R_{\rm sing}=\lambda^{-1}$ (dotted).
Also shown are the minimal horizon radius ${R}_{\rm min}$ 
and its asymptote (dotted).
}
\end{center}
\label{fig2}
\end{figure}

We conclude that the deformation of the horizon reflects 
itself in the location of the singularity. 
At the symmetric points $z=0$ and $z=1$, where the function $e^{C_{\rm H}}$
yields the minimal horizon radius ${R}_{\rm min}$, 
the distance $R_{\rm sing}$ of the singularity from the horizon 
is also minimal.
In particular, for large $\lambda$,
${R}_{\rm min} \sim \lambda^{-1}$ along with $R_{\rm sing}$
(see Fig.~3).
Extrapolating to the limit of infinite deformation, we infer that
the singularity touches the horizon at $z=0$ and $z=1$,
but is hidden well inside the horizon everywhere else.

In order to get further insight into the geometry of space 
in the NUBS interior,
we consider an isometric embedding of surfaces of constant 
Kretschmann scalar $K= R_{\mu\nu\rho\sigma} R^{\mu\nu\rho\sigma}$.
Thus, if such a surface has coordiantes $r_K(z),z,\theta_1,\theta_2,\varphi$ 
(at a fixed time), we define coordinates $X_i,Z$ by 
\begin{eqnarray}
dX_i dX^i + dZ^2 & = & 
e^{2\hB} \left(-\left(dr_K/dz\right)^2+1\right)dz^2 + e^{2\hC} d\Omega_{3}^2 
\nonumber \\
& = & dR^2 + dZ^2 + R^2d\Omega_{3}^2 \ , 
\end{eqnarray}
where the $X_i$ are expressed in terms of spherical coordinates 
in the second line.
Regarding $R$ and $Z$ as functions of $z$ then yields $R(z)= e^{\hC}$ and 
$$
Z(z) = L \! \int_{\frac{1}{2}}^z \! \! \sqrt{
e^{2\hB}\left(1-\left(dr_K/dz\right)^2 \right)
-e^{2\hC}\left(d\hC/dz\right)^2 
}\, dz' \, 
$$
where $\hB$ and $\hC$ are taken along the curve $(r_K(z),z)$,
and the length scale $L$ is reintroduced.

\begin{figure}[h!]
\begin{center}
\epsfysize=5cm
\mbox{\epsffile{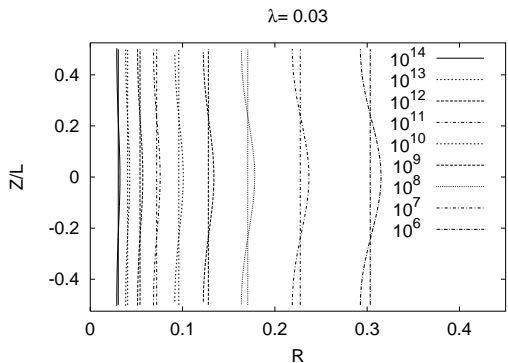}}
\caption{Isometric embedding of surfaces of constant Kretschmann scalar $K$
for NUBS with $\lambda=0.03$.
The straight lines correspond to
surfaces of constant $K$ of UBS with the same temperature.}
\end{center}
\end{figure}

We exhibit surfaces $Z(R)$ of constant Kretschmann scalar $K$
for NUBS with deformation $\lambda=0.03$ and $\lambda=0.47$
in Figs.~4 and 5, respectively \cite{footnote2}.
The limit of infinite $K$ corresponds to the singularity at $R=0$. 
For UBS, which are also shown, $R \sim K^{-\frac{1}{8}}$, independent of $Z$.
The NUBS surfaces, on the other hand,
show an oscillation about constant $R$ values,
and thus a $Z$ dependence of the power law. 
Here $R$ tends more slowly to zero for $Z=0$ 
and faster for $Z=\pm L_K/2$, 
where $L_K$ denotes the periodic length of $Z$.
Note, that $L_K$ increases with increasing $K$.
Clearly, the deformation of the horizon reflects itself also in the
isometric embeddings of the surfaces of constant $K$.

\begin{figure}[t!]
\begin{center}
\epsfysize=5cm
\mbox{\epsffile{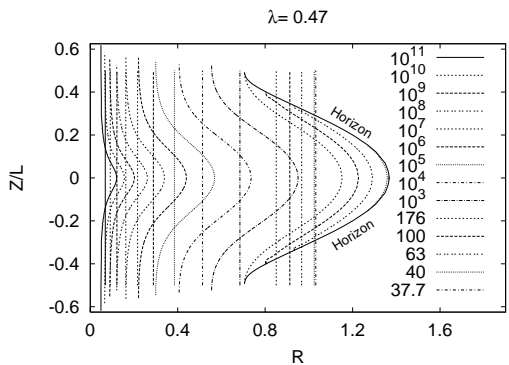}}
\caption{Same as Fig.~4 for $\lambda=0.47$. The horizon is also shown.}
\end{center}
\end{figure}

{\sl Outlook --}
Concerning the black hole -- black string transition
we have provided numerical evidence that at the merger point
the NUBS singularity touches the horizon at the waist of the strings.
What is still missing is the study of the interior of the caged
black holes, as they approach the merger point,
to see how their curvature and, in particular, their singularity evolve.
In $D=4$ such a study has been performed recently \cite{Frolov:2007xi},
in higher dimensions, however, this remains a major numerical challenge.

A GL instability arises also for rotating NUBS \cite{Kleihaus:2007dg},
and a similar instability occurs for
asymptotically flat rotating black holes in $D \ge 6$ dimensions,
which have a single angular momentum \cite{Emparan:2003sy}.
This instability then suggests, that as in the case
of black strings, a branch of rotating `pinched' (nonuniform) black holes 
should arise at the marginally stable solution \cite{Emparan:2003sy}.
Exploring this analogy further
leads to an intriguing phase diagram for rotating black holes,
where horizon topology changing transitions 
from rotating pinched black holes to black rings,
black saturns, and further configurations 
are expected to occur \cite{Emparan:2007wm}.

{\sl Acknowledgment --}
We gratefully acknowledge discussions with V.~Frolov, E.~Radu and A.~Shoom.
BK was supported by the German Aerospace Center.


\begin{thebibliography}{000}
\bibitem{footnote}
For the numerical construction of the solutions we have also employed
the parameterization
$
q(r,z)=e^{-4 \hat A}\ , \ \ w(r,z)=e^{-2(\hat B+\hat C)}\ ,
\ \ \gamma(r,z)=e^{3 \hat C + \hat A} \ .
$
\bibitem{footnote2}
Note, that close to the horizon the embedding is pseudo-Euclidean
with $dZ^2 \rightarrow -dZ^2$.
\end{thebibliography}
\end{document}